\documentclass[lettersize,journal]{IEEEtran}
\usepackage{amsmath,amsfonts}
\usepackage{algorithmic}
\usepackage{algorithm}
\usepackage{array}
\usepackage[caption=false,font=normalsize,labelfont=sf,textfont=sf]{subfig}
\usepackage{textcomp}
\usepackage{stfloats}
\usepackage{url}
\usepackage{verbatim}
\usepackage{graphicx}
\usepackage{cite}
\begin{document}

\title{Integrated Microwave Photonics True-Time Delay Signal Processor\\}

\author{Pablo Martínez-Carrasco, Tan Huy-Ho, Jose Capmany
\thanks{This work was supported by Huawei through contract YB20200065124. \textit{(Corresponding author: Pablo Martínez-Carrasco).}}
\thanks{Pablo Martínez-Carrasco and Jose Capmany are with the Photonics Research Labs, iTEAM Research Institute, Universitat Politècnica de València, Valencia, Spain (e-mail: pmarrom@iteam.upv.es; jcapmany@iteam.upv.es).\\
Tan Huy-Ho is with the Ottawa Wireless Advanced System Competency Centre, Huawei Technologies Canada Co.,Ottawa, Canada. (e-mail: tan.ho@huawei.com). }}

\markboth{January 2025}%
{Shell \MakeLowercase{\textit{et al.}}: A Sample Article Using IEEEtran.cls for IEEE Journals}
\maketitle

\begin{abstract}

Silicon photonic integrated circuits offer significant improvements in processing bandwidth, power efficiency, and low latency, addressing the needs of future microwave communication systems. Several successful applications have been demonstrated in this field; however, the focus is now shifting toward integrating these applications into single programmable photonic circuits. This approach not only reduces fabrication costs but also makes photonics more accessible for everyday use. This paper presents a scalable silicon-based signal processor with advanced functionalities, including high-speed arbitrary waveform generation,tunable bandwidth filtering and ultrabroadband beamforming. These results highlight improvements in both scale and performance, representing a significant step forward in large-scale, high-performance, multifunctional photonic systems.

\end{abstract}

\section{Introduction}
\label{sec:intro}
The rapid growth of data volumes in communications, AI training, and cloud computing demands efficient data processing. This data is usually stored as digital electrical information and transmitted as wireless radio frequency (RF) signals. However, modern communication systems and computational electronics may struggle to handle the increasing workload, making it essential to explore alternative processing techniques.
Microwave photonics (MWP) have emerged as promising solutions to these challenges. These systems harness the unique advantages of optics—such as flexibility, high bandwidth, and low power consumption—to process millimeter-wave signals directly in the optical domain \cite{MWP1,MWP2}. These techniques fit perfectly with the recent revolution in fabrication methods for integrated photonic circuits (PICs), which allow the creation of large, complex systems with exceptional precision \cite{photonics_future1,photonics_future2,photonics_future3}. This breakthrough has sparked significant interest across a wide range of research areas, including emerging computational paradigms such as quantum \cite{quantum1,quantum2} and neuromorphic computing \cite{neuro1,neuro2}, advanced sensing and spectroscopy systems \cite{sens1,sens2}, and the aforementioned optical signal processing technologies \cite{signal1,signal2}.

Integrated microwave photonic circuits offer flexible, broadband analog solutions that bridge the limitations of digital signal processors (DSPs), currently constrained to a few gigahertz. These circuits are designed to meet the growing demands of modern telecommunications, driven by advancements in Internet of Things , 5G, 6G, and cutting-edge radar systems \cite{fd1,fd2,fd3}. Many key RF functions have been successfully demonstrated using on-chip photonic signal processors, such as spectral filters \cite{app2}, RF phase shifters \cite{app1, app3}, integrators, differentiators \cite{app5}, pulse shapers and microwave generation \cite{app6,app7}, and beamformers \cite{app4}, all delivering exceptional performance. However, despite their exceptional performance, most of these circuits are still custom-designed for specific tasks. The future of this field lies in multifunctional integration, combining programmable functions onto a single chip to reduce fabrication costs and accelerate commercial implementation \cite{signal3, signal4}.

In this work, we propose a scalable design for a photonic signal processor chip based on optical true-time delay (TTD). This device aims to address three major applications of MWP: signal generation, filtering and beamforming. Our goal is to achieve a simple, universal, and versatile design while maintaining high performance to enhance future telecommunication networks. 

In Section \ref{sec:arch}, we present the proposed architecture, detailing its general implementation. This section also introduces the fabricated silicon chip used for the practical implementation and provides an explanation of the measurement setup and procedures. Moving forward, in Section \ref{sec:awg} we introduce the first application of the architecture, an ultrafast RF arbitrary waveform generator. Then, in Section \ref{sec:filter}  we demonstrate the second application of the architecture: a programmable filter with variable bandwidth. This section showcases a wide range of filter implementations, all achieved using the same device. In Section \ref{sec:beam} we show the results from the last functionality of the architecture, the beamforming network and finally, in Section \ref{sec:conc}, we provide a summary and conclusions of the work.

\begin{figure*}
\includegraphics[width=0.95\linewidth]{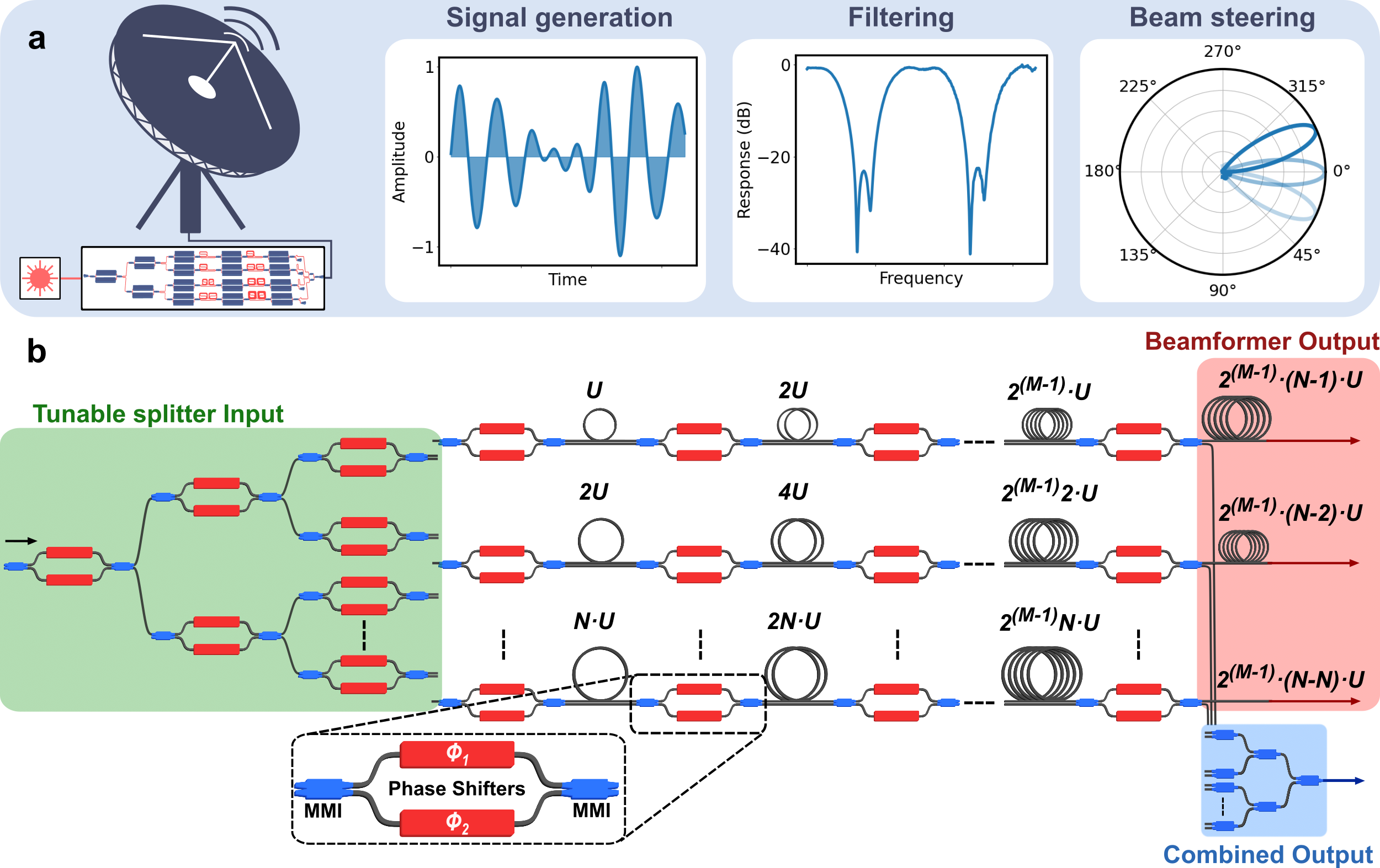} 
\caption{\textbf{(a)} Conceptual illustration of the proposed silicon photonic signal processor designed for multipurpose photonic processing applications.
\textbf{(b)} General schematic of the proposed architecture featuring $N$ lines and  $M$ delay stages. The tunable splitter tree at the input, highlighted in green, distributes light across the lines. MZIs guide the light through the delay stages, finishing at the backend, where they are used for switching between the two possible outputs: the beamformer output (highlighted in red) and the combined output (highlighted in blue).}
\label{fig:architec}
\end{figure*}

\section{\label{sec:arch}Architecture Design}

True-time delay is a widely used technique in microwave photonics for implementing beamforming networks that are free from beam squint. There are two main approaches to realizing TTD on integrated devices: microring resonators, which offer continuous delays but are sensitive to thermal fluctuations and fabrication errors, and switched delay lines, which are more robust and provide broader bandwidth, though they are limited to a discrete set of delay positions. For our design, we opted for the latter, aiming to maximize the number of achievable delay settings while ensuring high tolerance to deviations and imperfections. 
The architecture is closely related to our recent work, initially designed as a beamforming network \cite{beam}. In that work, we introduced a novel beamformer design that minimized the number of delays while maximizing resolution, optimizing resource efficiency. The objective is to preserve the high-resolution, ultra-broadband performance of the beamformer while integrating additional functionalities, such as filtering and arbitrary waveform generation (AWG). This is accomplished by incorporating a combined output at the back-end of the structure, formed by Multi-Mode Interferometers (MMIs).

The general scheme of the proposal is depicted in Fig. \ref{fig:architec}. The main building block of this design is the tunable coupler, implemented here as Mach-Zehnder Interferometers (MZI) formed by two 2x2 MMIs with a phase shifter section between them. In this work, the phase shifter operates using the thermo-optic effect, but for faster applications, the electro-optic effect can be considered. The general transfer function of a MZI can be expressed as follows:

\begin{equation}
{\scriptstyle
    H_{MZI}(\phi_{1},\phi_{2})
    = 
    ie^{i(\phi_{1}+\phi_{2})/2}
    \begin{bmatrix} 
     \sin{\left(\dfrac{\phi_{1}-\phi_{2}}{2}\right)} & \cos{\left(\dfrac{\phi_{1}-\phi_{2}}{2}\right)} \\
     \cos{\left(\dfrac{\phi_{1}-\phi_{2}}{2}\right)} & -\sin{\left(\dfrac{\phi_{1}-\phi_{2}}{2}\right)} \\
    \end{bmatrix}}
\label{eq:mzi}
\end{equation}

By adjusting the phase shifts in the segment between the MMIs, $\phi_{1}$ and $\phi_{2}$, we can adjust the overall phase shift of the optical carrier at the output, which is highly beneficial for filter synthesis. More importantly, we can simultaneously vary the power splitting ratio between the arms of the MZI, a feature that enables the creation of diverse beam patterns or the selection of different windows funtions. Generally, when operating the core of the device, the tunable couplers are typically configured in either the bar or cross state. This configuration is well-suited for representing binary notation, with each state commonly encoded as a bit word.

At the front-end of the structure, there is a tunable splitter tree which depends on the number of total lines, $N$. This setup allows for customized power distribution to each line, rather than a uniform distribution across all lines, without introducing additional losses. Afterwards, connected to each output of the splitter tree, there is a delay line with $M$ delay stages or bits, where the delay values increase exponentially in powers of 2 for each bit with respect to a fundamental delay, $U$. As a specific case of this architecture, the delay value also scales with the position of the line. For instance, the delay value for the second line is doubled compared to the first line, and for the $N^{th}$ line, it is multiplied by $N$. This results in $2^{M}$ unique combinations of delays between lines at the output of the final MZI, with the total delay between antennas given by $\Delta T = BU$, where $B$ is the selected bit word ranging from $0$ to $(2^{M}-1)$. 

The final MZI manages the switch between the two distinct outputs of the architecture: the beamformer and the combined output. While the combined output simply transfers the original delays achieved in the device's core to the common combiner formed by MMIs, the waveguide of the beamformer output introduces an additional delay depending on the line position, given by $\epsilon_{n}=(N-n) 2^{(M-1)}U$. This adjusts the range of delays between antennas while preserving all combinations, shifting the value of  $\Delta T$ from $[0, (2^{M}-1)U]$ to $[-2^{M-1}U, (2^{M-1}-1)U]$. This adjustment allows the beamformer to aim without beam squint effects across both positive and negative angles, maintaining constant delays between antennas in both directions of the array while maximizing the number of delay stages to achieve the highest possible resolution.

\subsection{Experimental demonstration}

For the experimental demonstration, we included four delay lines, each with two delay stages (2 bits of resolution), fed by a tunable splitter tree, with outputs connected using a fixed combiner consisting of three 2x1 MMIs. The fundamental delay, $U$, was designed to be 10 picoseconds, slightly larger than the width of the pulse-shaped laser available in our laboratory, which was intended for measuring the AWG. This consideration is crucial to prevent the laser pulses from overlapping at the combiner. For this measurement, we used high-speed fiber-pulsed source that produced a train of Gaussian pulses with a width of 4.4 ps and a repetition frequency of 5 GHz. The repetition rate was chosen to be sufficiently low to ensure that there is no overlap between pulses at the chip's output.

\begin{figure}[ht]
\centering
\includegraphics[width=0.99\linewidth]{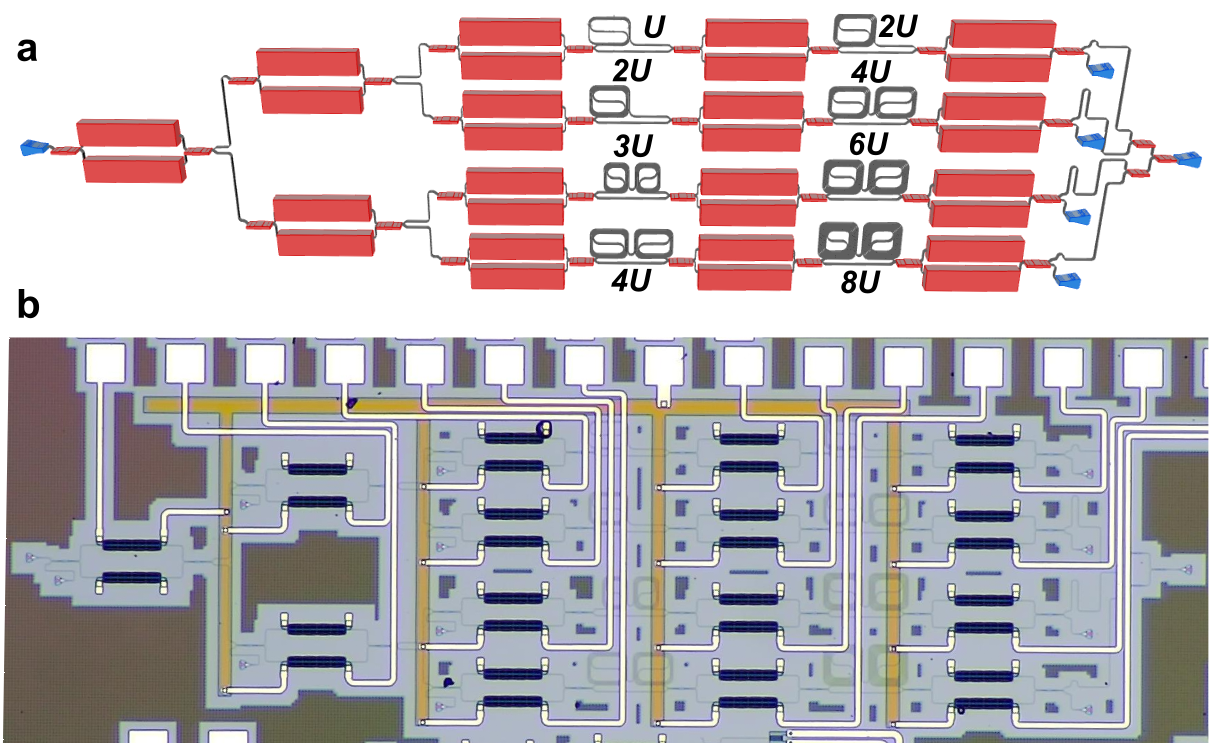} 
\caption{\textbf{(a)} Layout of the experimental demonstration chip. \textbf{(b)} micrograph of the fabricated device on a SOI platform (4mm x 1.2mm).}
\label{fig:chip}
\end{figure}

We designed a photonic integrated circuit that was manufactured by Advanced Micro Foundry on a Silicon on Insulator platform. The layout and a micrograph of the chip are presented in Figure \ref{fig:chip}. The chip was produced using an SOI wafer with a 220 nm slab thickness, with 500 nm single-mode waveguides that were defined through deep ultraviolet lithography (193 nm). For the  sections containing phase shifters, a thin heater layer of 120 nm TiN was deposited over the waveguide giving as a result MZIs with a measured power consumption of only $1.35mW/\pi$. The thermo-optic phase shifters were controlled electronically using Qontrol programmable power sources, managed via software.

The optical input and output of the chip were achieved through vertical coupling using grating couplers, with coupling losses of 3 dB. These losses were normalized during the measurements, isolating propagation and insertion losses as the primary loss mechanisms within the chip. The propagation losses were approximately 1 dB/cm, while the insertion losses for each MZI were around 0.4 dB.   

Time delay and transmission measurements were conducted using a Vector Network Analyzer (VNA) paired with a Lightwave component module. This module integrates an optical modulator and photodetector, enabling direct electrical-to-optical up-conversion within a single device. 

\section{AWG application}
\label{sec:awg}

\begin{figure*}[ht]
\includegraphics[width=0.99\linewidth]{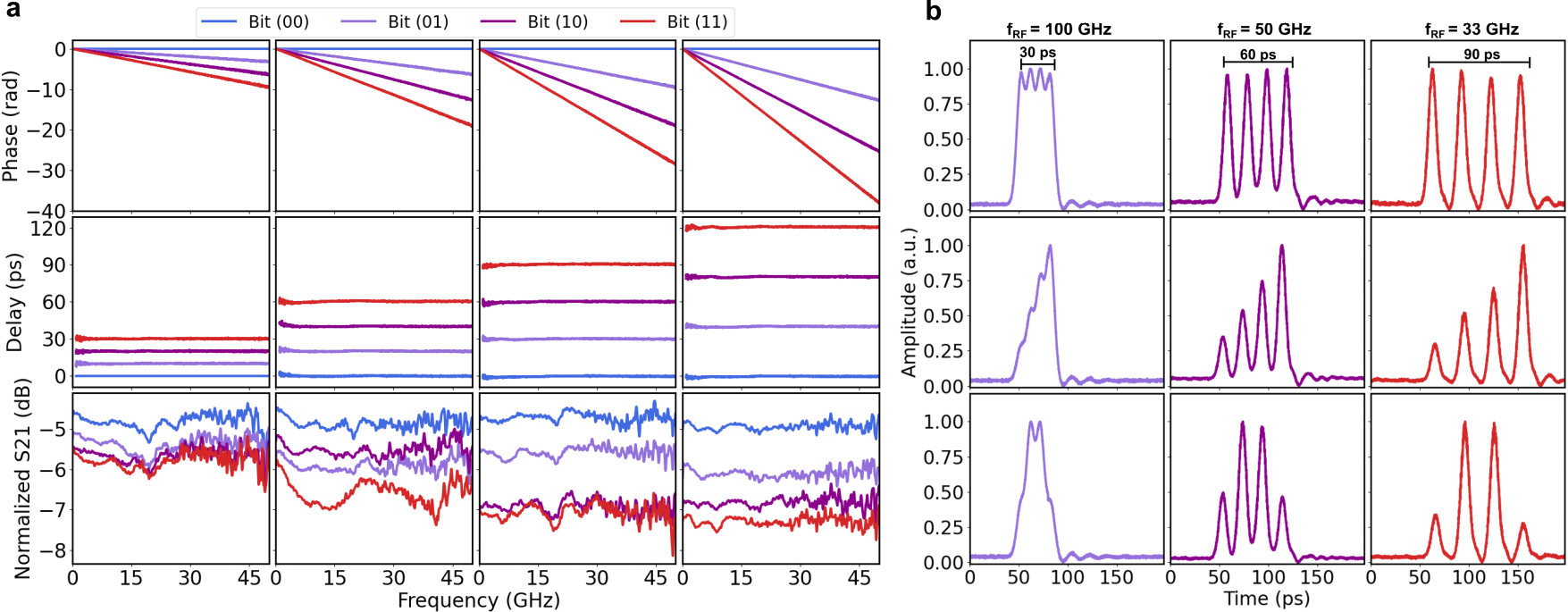} 
\caption{\textbf{(a)} Results from the RF characterization of the delay lines. Each column represents measurements from the first to the last delay line in the architecture, while the colored lines indicate the different bit configurations of all the lines. \textbf{(b)}  Measured waveforms for different time separations between peaks and shapes, including rectangular, sawtooth, and Gaussian profiles.}
\label{fig:delays}
\end{figure*}

Ultrafast optical and radio-frequency waveforms with bandwidths reaching tens to hundreds of gigahertz have potential applications in numerous fields, including high-speed optical communications, radar, microwave imaging, and instrumentation test measurements \cite{awg5}. However, generating high-frequency (e.g., 20–80 GHz) and ultrabroad-bandwidth arbitrary waveforms through electronic means is challenging due to the limitations of digital-to-analog converter technology and significant timing jitter \cite{awg1,awg2}. Because of that, AWG has become a focal area in microwave photonics due to the ability of optical techniques to produce high-frequency and large-bandwidth signals. The most common methods for generating arbitrary waveforms in photonics rely on bulky frequency-to-time mapping techniques. This process involves adjusting a broadband source using a spectral filter and then passing the broadband signal through a dispersive element. In the dispersive element, different wavelengths are delayed relative to each other, resulting in their arrival at a photodetector at different times, thus creating the desired waveform \cite{awg3,awg4}. 

Here, we demonstrate a compact device capable of generating arbitrary waveforms using a temporal pulse shaping system synthesized in the time domain. This approach eliminates the need for external dispersive elements while keeping control over both the shape and frequency of the generated waveform. This is accomplished through the generation of delayed and scaled replicas of short input optical pulses in the optical domain, followed by their recombination and scaling to produce various RF wavepackets. By adjusting the intertap delay, we can expand or contract the replicas from the input signal. The splitter tree enables the application of different weights, facilitating the creation of complex waveforms. 

First, we characterized the processor by precisely controlling the optical coupling of all the MZIs in the design with respect to the applied electrical power. After completing the MZI characterization, we proceeded with delay measurements using the VNA to verify that all delays were accurately implemented. Figure \ref{fig:delays}(a) presents the results obtained from measuring the RF transmission, phase and time responses. The time and delay measurements were highly satisfactory, aligning perfectly with the theoretical predictions.
The RF transmission spectra exhibit some noise, which may be attributed to the finite extinction ratio of the MZIs, thermal crosstalk, and electrical variations caused by the use of multi-contact probes. However, these variations are relatively minor, remaining within 1 dB, and are expected to improve significantly in fully packaged devices. On the other hand, the losses progressively increase from one line to the next due to propagation losses, which are inherent and unavoidable. This effect, however, was accounted for and mitigated using the tunable splitter tree, which compensates for the higher losses by distributing more optical power to the affected lines.

In our proof-of-concept design, the total number of delay combinations is four, as summarized in Table \ref{tab:delays}.  The first bit word is not relevant for this application, as the lack of delay between pulse replicas causes them to arrive simultaneously at the combiner, avoiding any waveform generation.

\renewcommand{\arraystretch}{1.3} 
\setlength{\tabcolsep}{7pt} 
\begin{table}[ht]
\centering
\begin{tabular}{c || c c c c|}
\cline{2-5}
     & \textbf{Bit (00)} & \textbf{Bit (01)} & \textbf{Bit (10)} & \textbf{Bit (11)} \\ \hline \hline
\multicolumn{1}{|c||}{\textbf{Line 1}} & 0 & U & 2U & 3U   \\ 
\multicolumn{1}{|c||}{\textbf{Line 2}} & 0 & 2U & 4U & 6U  \\ 
\multicolumn{1}{|c||}{\textbf{Line 3}} & 0 & 3U & 6U & 9U  \\ 
\multicolumn{1}{|c||}{\textbf{Line 4}} & 0 & 4U & 8U & 12U \\ \hline
\end{tabular}
\caption{Value of the delay applied on each line for all the bit words.}
\label{tab:delays}
\end{table}

Figure \ref{fig:delays}(b) displays the measured optical waveform on our chip before photodetection. The separation between the peaks is inversely related to the resulting radio frequency signal generated on a high-speed photodetector. Using our fabricated device, we can produce optical replicas with three distinct separations, corresponding to generated RF center frequencies of 33 GHz, 50 GHz, and 100 GHz, depending on the selected bit word: (11), (10), and (01), respectively.

\section{Filter application}
\label{sec:filter}

Microwave filters are crucial components in RF links and are among the applications that benefit most from photonic integration, as photonics enhances these systems through RF-to-optical frequency upconversion and flexible optical-domain filtering. Recent advancements in integrated optics present significant opportunities for innovative filter designs, including both infinite impulse response (IIR) and finite impulse response (FIR) filters, each offering distinct characteristics that make them well-suited for specific performance requirements and applications \cite{filt1}.

In this work, we demonstrate two types of FIR filters achievable with our proposed architecture. The first type, multi-tapped delay filters, also known as transverse filters, are inspired by the digital filter concept commonly used in RF systems and represent the most traditional approach for implementing MWP filters. In a transverse filter, the input signal undergoes discrete sampling, followed by delaying and weighting, before being summed. All these steps can be implemented using optics, creating a transfer function described by:

\begin{equation}
    H(\omega) = \sum^{N-1}_{n=0}a_{n}e^{-i\alpha_{n}}e^{-in\omega\Delta T}
\end{equation}

Where $a_{n}$ is the weight of the sampled signal at the $n$-th tap, $\alpha_{n}$ is the carrier phase shift of the $n$-th sample, and $\Delta T$ is the intertap signal time delay. The frequency selectivity of the filter is primarily governed by the number of effective taps, the inter-tap delay sets the frequency periodicity with the free spectral range (FSR) given by $1/\Delta T$, and the phase coordination establishes the spectral positions of both the filter’s passbands and stopbands \cite{filt2}.

\begin{figure}[ht]
\centering
\includegraphics[width=0.99\linewidth]{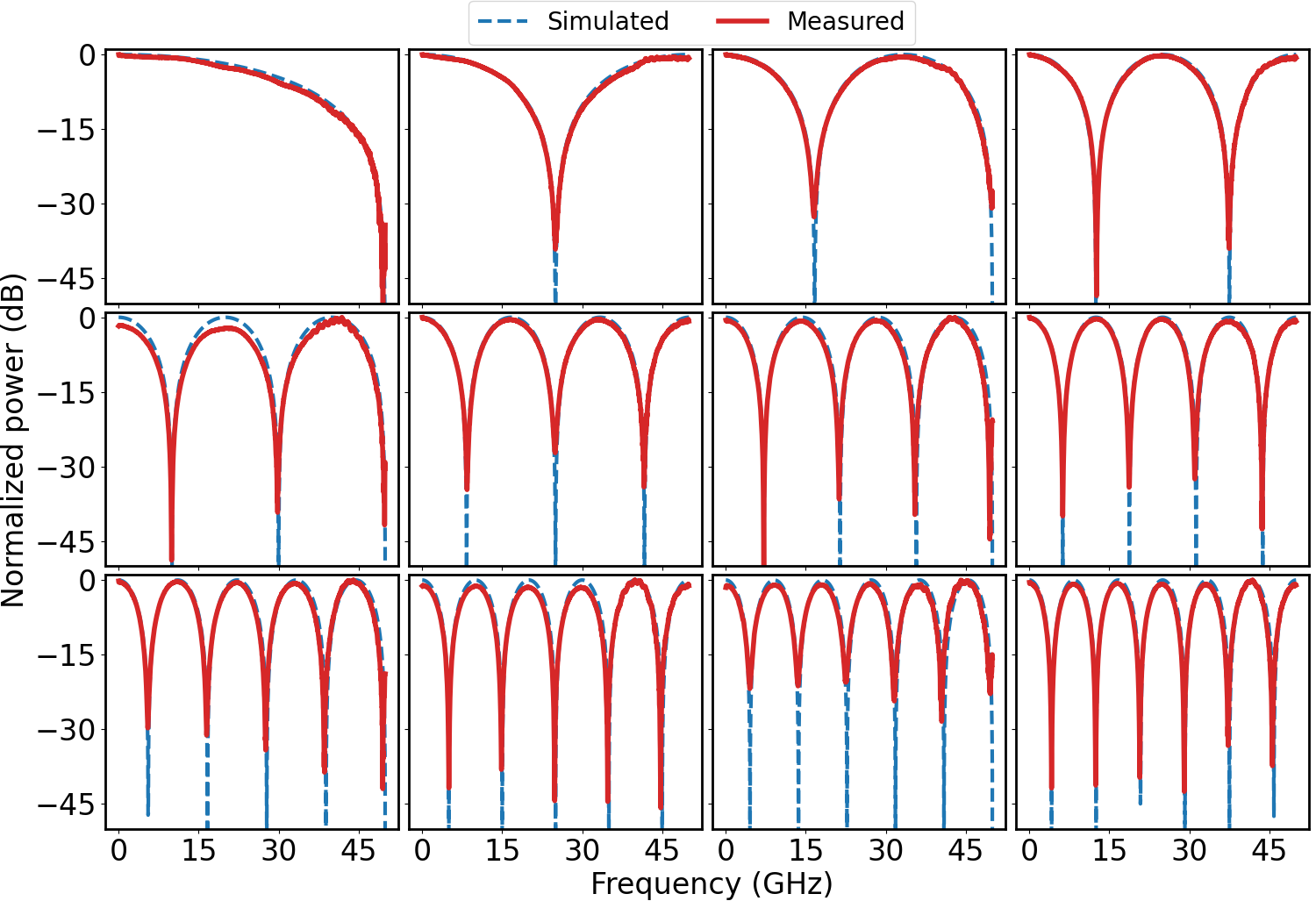} 
\caption{Comparison between simulated and measured results for all 12 two-tap filter combinations, arranged in order of increasing $\Delta T$ values, from the smallest to the largest.}
\label{fig:2Tap}
\end{figure}

Thanks to the flexible splitter tree at the system's front end, we can select between different feeding modes, simultaneously driving two, three, or all four lines, enabling a wide variety of filter shapes and bandwidth combinations. When using only two lines, it is possible to configure up to 12 distinct intertap delays between them, ranging from $U$ to $12U$. Consequently, the free spectral range varies, resulting in a tunable notch filter with an FSR ranging from 100 GHz when  $\Delta T = U$ to 8.3 GHz when $\Delta T = 12U$. Figure \ref{fig:2Tap} shows all the different filter combinations of two taps.

\begin{figure}[ht]
\centering
\includegraphics[width=0.99\linewidth]{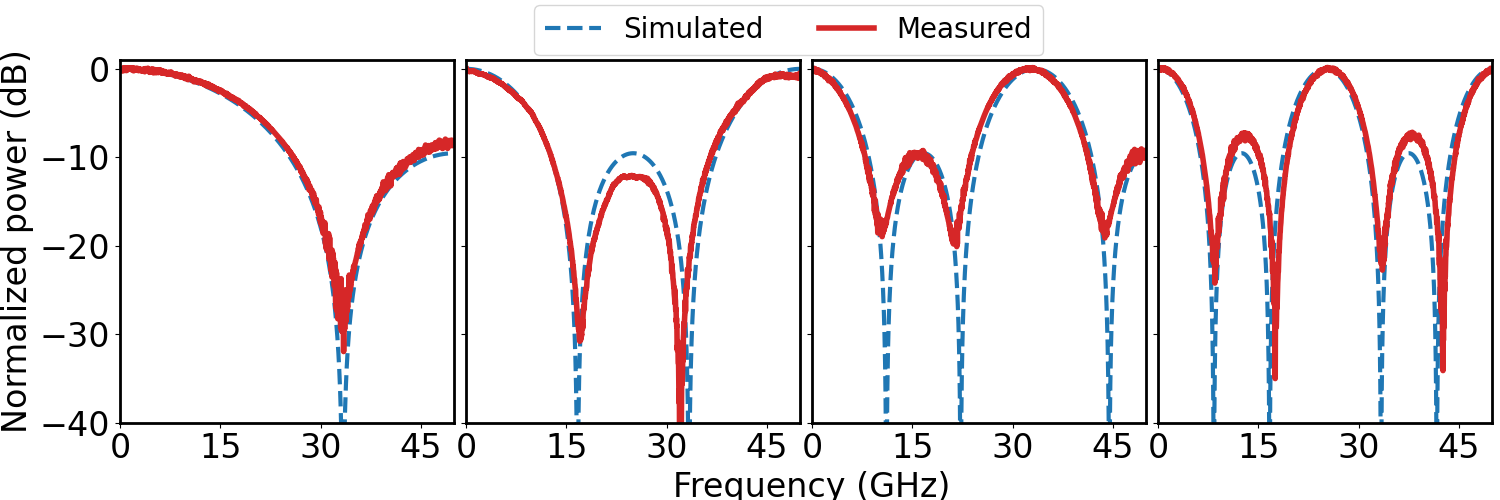} 
\caption{Comparison of simulated and measured filter responses with three lines in use.}
\label{fig:3Tap}
\end{figure}

Similarly, by feeding three lines, we obtain an additional tap for the filter. In this configuration, there are only four possible delay combinations, ranging from $\Delta T = U$ to $\Delta T = 4U$. As a result, the filter's response transitions from a notch filter to a typical transverse filter, characterized by variable bandwidth. Specifically, we measured 3 dB bandwidths of 32.1 GHz, 14.8 GHz, 8.9 GHz, and 8.45 GHz, corresponding to FSRs of 100 GHz, 50 GHz, 33.3 GHz, and 25 GHz, respectively. The results from the measurements are shown in Fig. \ref{fig:3Tap}.

\begin{figure}[ht]
\centering
\includegraphics[width=0.99\linewidth]{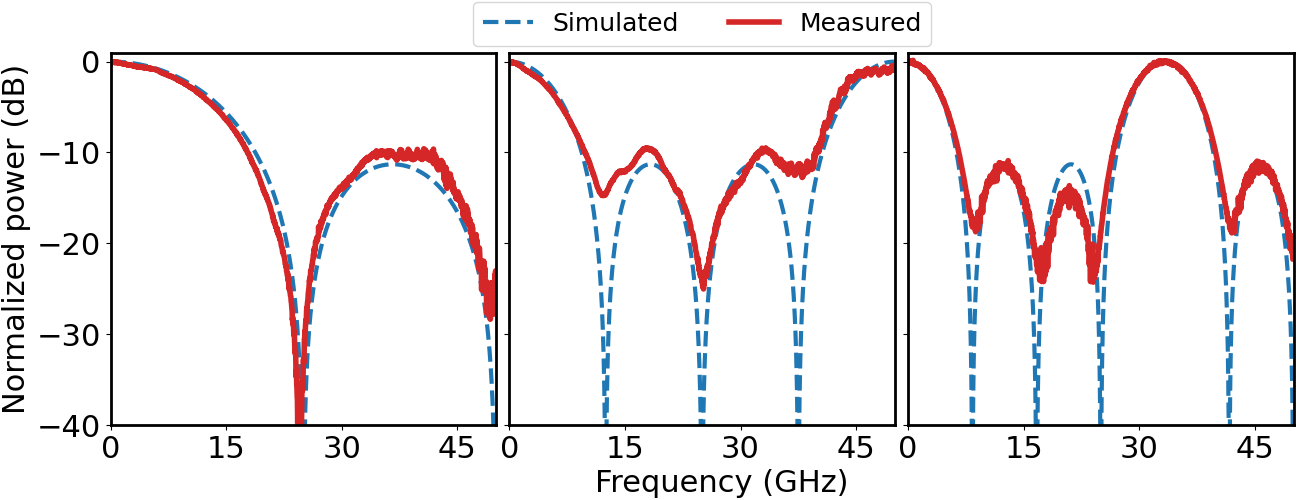} 
\caption{Measured filter responses with all lines fed simultaneously, showing bandwidths of 21.5 GHz, 10.3 GHz, and 7.8 GHz.}
\label{fig:4Tap}
\end{figure}

Finally, all the lines can be used simultaneously, resulting in three possible intertap delays. This results in narrower filter bandwidths while preserving the same FSR. Figure \ref{fig:4Tap} depicts the results for the four-tap measured filters.

\begin{figure}[ht]
\centering
\includegraphics[width=0.95\linewidth]{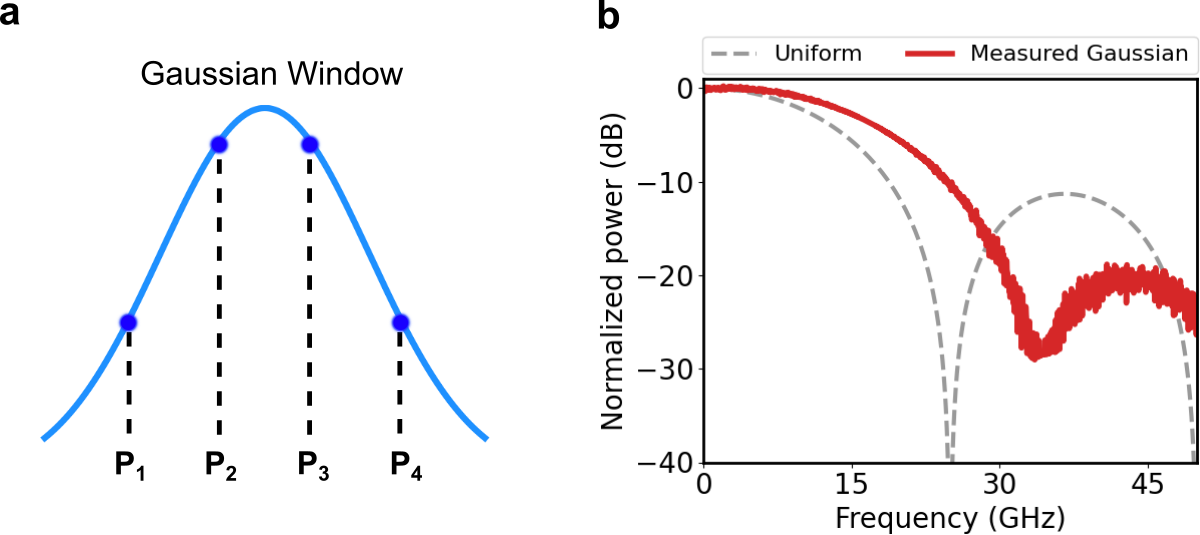} 
\caption{\textbf{(a)} Illustration of a Gaussian window applied to a four-tap filter, where the optical power, $P$, distributed across each line follows the profile of a Gaussian function. \textbf{(b)} Measurement results for the filter with an intertap delay $\Delta T = U$, comparing the performance of a Gaussian window with uniform power distribution.}
\label{fig:gauss}
\end{figure}

Additionally, the tunable splitter tree at the input of the architecture allows for flexible distribution of optical power across all the lines. This capability enables the implementation of window functions, a common tool in signal processing, within the synthesized filters. By applying window functions, we can enhance the Main Lobe to Secondary Lobe Ratio (MLSLR) of the filter, as illustrated in Fig.\ref{fig:gauss}. However, this improvement comes with a slight trade-off, as it increases the bandwidth of the low-pass band.

The second type of filters that can be implemented with this approach are two-stage lattice filters. In this configuration, the light is routed through a single line, and by adjusting the coupling strength of each MZI (rather than strictly using the bar or cross states as in the previous case), it is possible to achieve different band-pass filter responses.
The general transfer function for a filter implemented on the $N^{th}$ line, with $M$ delay stages (corresponding to $M+1$ MZIs), is given by:
\begin{equation}
    H(\omega)= H_{MZI_{M+1}} \prod_{m=1}^M 
    \begin{bmatrix} 
     e^{-i\omega 2^{(m-1)}NU} & 0 \\
     0 & 1 \\
    \end{bmatrix}
    H_{MZI_{m}} 
\label{eq:lattice}
\end{equation} 

It is straightforward to observe that for identical coupling configurations of the MZIs across different lines, the only distinction in their transfer functions lies in their FSR, associated with the value of the smallest delay in the line, given by $FSR = 1/(NU)$. The common phase can be ignored because it will be absorbed by the photodetector, simplifying the analysis. The transfer function of the MZIs from Eq.(\ref{eq:mzi}) can be simplified to that of a tunable coupler using the relations: $\sin{\left(\Delta\phi_{m}/2\right)} = \sqrt{1-\kappa_{m}}$ and $\cos{\left(\Delta\phi_{m}/2\right)} = \sqrt{\kappa_{m}}$ , depending only on the coupling coefficient $\kappa$.

\begin{figure}[ht]
\centering
\includegraphics[width=0.99\linewidth]{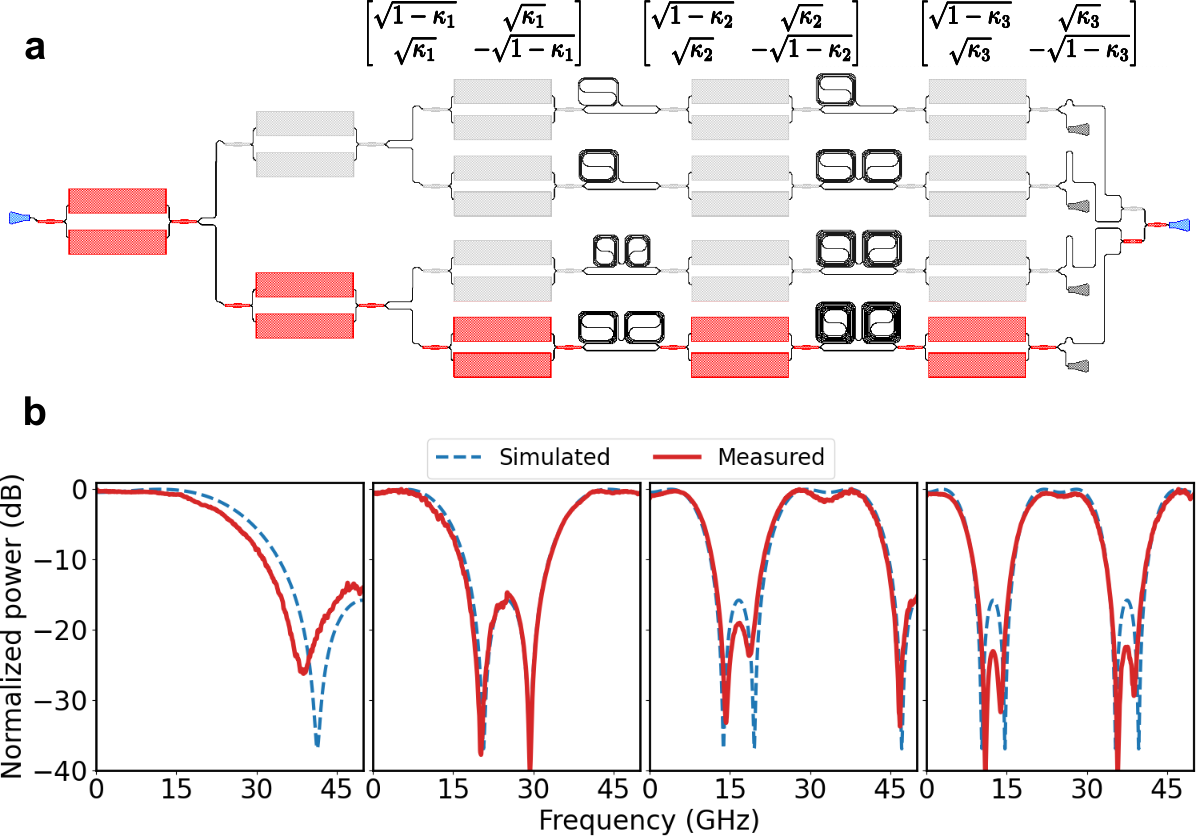} 
\caption{\textbf{(a)} Schematic illustrating the method for driving the fabricated chip to synthesize a lattice filter using the fourth line. \textbf{(b)} Results from implementing the same filter across all lines using a consistent set of coupling coefficients: $\kappa_{1} = 0.8$, $\kappa_{2} = 0.7$ and $\kappa_{3} = 0.5$.}
\label{fig:lattice}
\end{figure}

This concept is illustrated in the upper part of Figure \ref{fig:lattice}, which also demonstrates the implementation of one of these lattice filters in the fourth line of our device. Additionally, the figure presents measurement results using the same set of coupling coefficients applied across all the lines, demonstrating that while the filter shape remains consistent, the FSR varies.

\section{Beamforming application}
\label{sec:beam}

The final demonstrated application is the operation of the device as an optical beamforming network. This functionality was extensively covered in our previous work \cite{beam}; therefore, additional compensation delays were omitted in this design to optimize space in the proof-of-concept device. Instead, we prioritized demonstrating the other applications, as the added waveguide length used for beam reorientation does not introduce measurement challenges or affect the results.
Thus, for beam reconstruction we assume the presence of additional delays, as described in Section \ref{sec:arch}.

To evaluate the performance of the integrated beamformer, we calculate the far-field pattern of the emitted beams using the delay measurements from each channel and the array factor equation for uniform feeding, which is expressed as follows:

\begin{equation}
    F(\omega,\theta) = \sum_{n = 0}^{N-1}{\dfrac{1}{N}e^{-i\omega\left(\dfrac{d}{c}n\sin(\theta) - T_{n} \right)}}
\end{equation}

where $T_{n}$ represents the total delay at the output of the $n^{th}$ line and $d$ the spacing between adjacent antenna elements. In this case, $d$ was chosen to be 10 mm, ensuring that the resulting four steering angles range from -35º to 18º, while maintaining frequency independence for a maximum emission frequency of 16.5 GHz.

\begin{figure}[ht]
\centering
\includegraphics[width=0.99\linewidth]{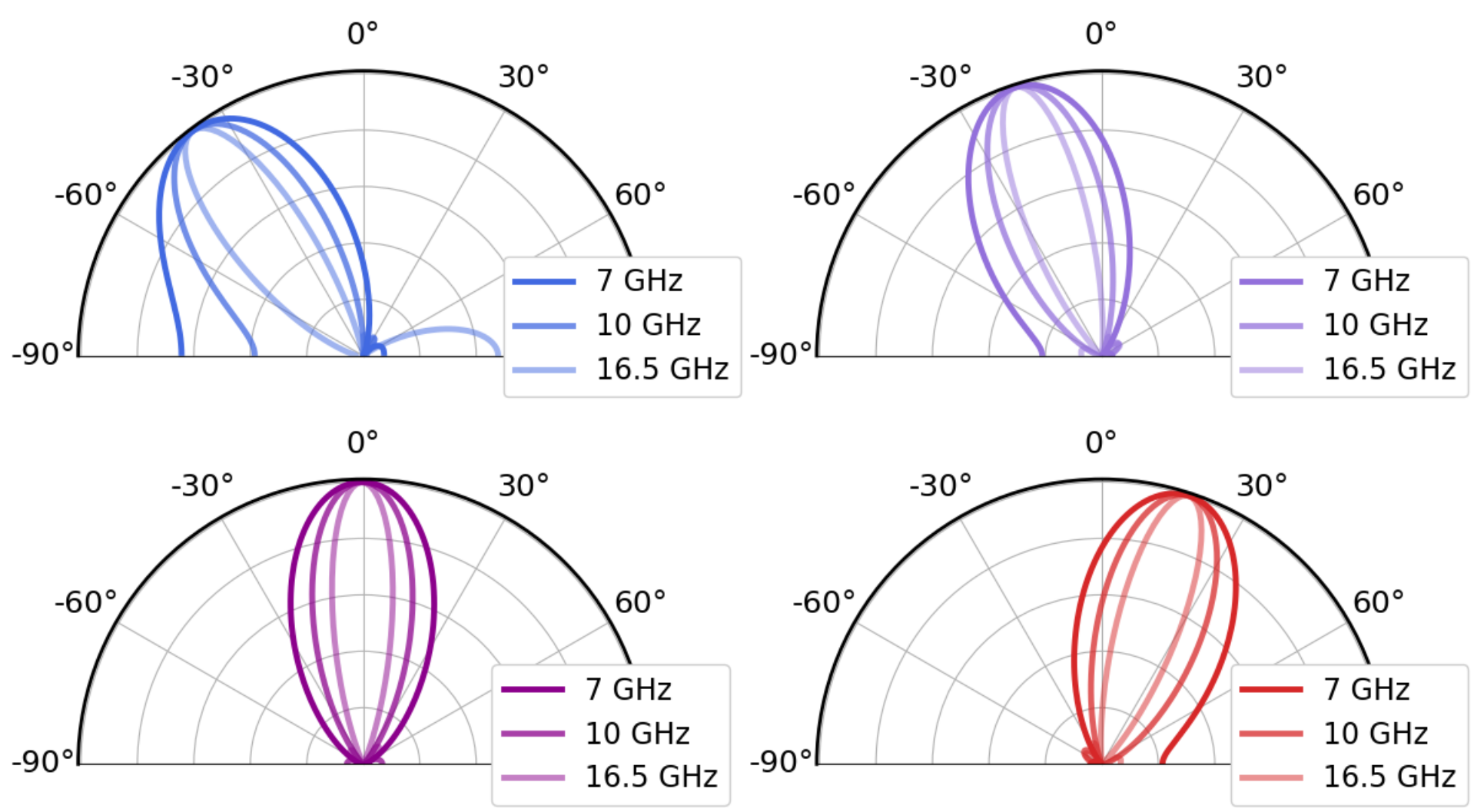} 
\caption{Simulated radiation patterns for all the bit configurations based on the measured delay values at emission frequencies of 7 GHz, 10 GHz, and 16.5 GHz.}
\label{fig:beams}
\end{figure}

Figure \ref{fig:beams} shows the beampatterns for a range of frequencies from 7 GHz to 16.5 GHz. Since the array consists of only four antennas, operating at lower frequencies is not recommended, as the beam width increases, reducing the array's directivity. On the other hand, operating at frequencies higher than 16.5 GHz is also not recommended, as it may lead to the emergence of undesired radiation lobes at the largest pointing angles.

\section{Conclusion}
\label{sec:conc}

In this paper, we have demonstrated a multifunctional and programmable architecture that operates as an arbitrary waveform generator, a tunable filter with adjustable bandwidth, and a broadband RF beamforming network.
These functionalities were successfully demonstrated on a fabricated silicon chip, achieving RF signal generation with various waveforms up to 100 GHz, a reconfigurable filter capable of operating as either a notch or band-pass filter with a wide range of FSRs and bandwidths, and broadband optical beamforming from 7 GHz to 16.5 GHz with a scanning range from -35º to 18º.

The results obtained are very satisfactory given the size of the test sample; however, the design is easily scalable by incorporating additional delay stages or lines, significantly expanding the range of achievable configurations across all functionalities while maintaining a small footprint.  This opens the door to potential implementation in future base stations or wireless communication systems, contributing to the advancement of next-generation communications.

\section{acknowledgments}
The authors wish to acknowledge the financial support of Huawei through contract YB20200065124.

\section*{Author Declarations }

The authors have no conflicts to disclose.

\section*{Data Availability Statement}

The data that support the findings of this study are available from the corresponding author upon reasonable request.

\end{document}